\documentclass[aip,pop,reprint]{revtex4-1}

\usepackage{amsmath}
\usepackage{upgreek}
\usepackage{graphicx}

\graphicspath{{./figs/}}

\begin{document}

\title{Gamma-ray generation in ultrahigh-intensity laser-foil
interactions}
\date{\today}
\author{E.~N.~Nerush}
\email{nerush@appl.sci-nnov.ru}
\author{I.~Yu.~Kostyukov}
\affiliation{Institute of Applied Physics of the Russian Academy of
Sciences, 46 Ulyanov St., 603950 Nizhny Novgorod, Russia}
\affiliation{University of Nizhny Novgorod, 23 Gagarin Avenue, Nizhny
Novgorod 603950, Russia}
\author{L.~Ji}
\affiliation{Heinrich-Heine-Universitat Dusseldorf, 40225 Dusseldorf,
Germany}
\author{A.~Pukhov}
\affiliation{Heinrich-Heine-Universitat Dusseldorf, 40225 Dusseldorf,
Germany}
\affiliation{University of Nizhny Novgorod, 23 Gagarin Avenue, Nizhny
Novgorod 603950, Russia}
%\pacs{pac}
%\keywords{kw}

\begin{abstract}

Incoherent photon emission by ultrarelativistic electrons in the
normal incidence of a laser pulse on a foil is investigated by means
of three-dimensional numerical simulations in the range of intensities
$2 \times 10^{21} \text{--} 2 \times 10^{25} \text{ W} \,
\text{cm}^{-2}$ and electron densities $2 \times 10^{22} \text{--} 1
\times 10^{24} \text{ cm}^{-3}$.  We focus on properties of the
resulting synchrotron radiation, such as its overall energy,
directivity of the radiation pattern and slope of the energy spectrum.
Regimes of laser-foil interactions are studied in the framework of a
simple analytical model.  The laser-plasma parameters for efficient
gamma-ray generation are found and revealed to be close to
the parameters for relativistic foil motion.  It is shown that in
the case of oblique incidence of a $3 \text{ PW}$, $10 \text{ fs}$ laser
pulse on a thin foil about $10^{8} \text{ photons}/0.1\% \text{
bandwidth}$ are produced at the energy level of $1 \text{ MeV}$ that
significantly exceeds performance of the modern Compton gamma-ray
sources. Various applications of the gamma-ray bunches are discussed. 

\end{abstract}

\maketitle

\section{\label{sec:intro}Introduction}

Nowadays gamma-ray sources are widely used in numerous applications
from medicine to nuclear physics.  Such sources can be based not only
on the radioactivity, bremsstrahlung~\cite{Wagner05} and synchrotron
emission in magnetic field~\cite{Bilderback05} but also on a Compton
scattering of laser light on electron beams of conventional
accelerators~\cite{Weller09,Gibson10,Albert10}.  The recent progress
in laser technologies and laser-plasma electron acceleration opens the
opportunity to eliminate a conventional electron accelerator from
the scheme and develop a compact all-optical Compton source of hard
photons~\cite{Ta12,*Powers14} or a laser-plasma synchrotron
source~\cite{Nerush07}.  Laser-produced hot electrons can also
generate bremsstrahlung
photons~\cite{Kmetec92,*Norreys99,*Hatchett00}, however, the
efficiency of this generation mechanism is not high due to a low value
of the bremsstrahlung cross-section. Nevertheless, the abundant
positron production via decay of bremsstrahlung photons in thick
high-Z targets is demonstrated~\cite{Chen09}.

Another mechanism of gamma-ray production is the nonlinear Compton
scattering induced directly by laser-driven plasma electrons.  For a
laser field of ultrarelativistic intensity the nonlinear Compton
scattering is equivalent to synchrotron
radiation~\cite{Nikishov64,*Ritus85}, i.e. ultrarelativistic
plasma electrons emit gamma-rays during travelling along curved
trajectories. The resulting radiation losses take away a considerable
part of the electron energy and significantly affect electron motion
if the laser intensity becomes greater than $10^{23} \text{
W} \, \text{cm}^{-2}$ (for optical wavelengths), the intensity of
the so-called radiation-dominated regime~\cite{Bulanov04,*Esarey93}.

Recent results on the generation of laser pulses with high intensity
(up to $2\times 10^{22} \text{ W}\, \text{cm}^{-2}$, see
Ref.~\onlinecite{Yanovsky08}) and a number of proposals for
multipetawatt and exawatt laser facilities~\cite{Mourou06,*Di12}
stimulate theoretical study of laser-plasma synchrotron sources.
E.g., generation of gamma-ray bunches in the interaction of a laser
pulse with a tailored overcritical density target is investigated on
the basis of theoretical analysis and 2D particle-in-cell simulations
with the radiation friction force incorporated~\cite{Nakamura12}.  The
theoretical analysis is done for a circularly polarized laser pulse
propagating in an underdense plasma. The obtained analytical results
agrees qualitatively with the results of simulations for
linear polarization.  More complex plasma dynamics than it is assumed
in Ref.~\onlinecite{Nakamura12} is revealed in
Ref.~\onlinecite{Brady12} for an intense laser pulse interacting with
a relativistically underdense plasma. The 2D
simulation shows that a great portion of plasma electrons moves
towards the laser pulse during a noticeable part of an optical cycle.
At this time interval the intensity of the synchrotron radiation emitted
by the electrons peaks, that leads to generation
of a $90^\circ$-wide gamma-ray radiation pattern directed towards the
laser pulse. Such back-and-forth electron motion (including the case of opaque
plasmas) is a key element of the high-harmonic generation
theory~\cite{Brugge10,*Gonoskov11,*Sanz12,*Bulanov13}, where a coherent part of
the synchrotron radiation from laser-irradiated solids is
investigated.  It is interesting to note that in
Ref.~\onlinecite{Ridgers12} a 2D simulation of a laser pulse
interacting with a relativistically opaque plasma reveals mostly
forward-directed gamma-ray radiation pattern.

Laser-plasma interactions potentially become even more complicated
at the intensity of $10^{24} \text{ W} \, \text{cm}^{-2}$, when
avalanche-like generation of gamma-quanta and electron-positron
($e^+e^-$) pairs can occur~\cite{Bell08, Fedotov10} due to consecutive
events of nonlinear Compton scattering and Breit--Wheeler process. The
produced $e^+e^-$ plasma can noticeably affect the interaction
process~\cite{Ridgers13} and can even cause a significant absorption of
the laser field~\cite{Fedotov10,Nerush11}. In the latter case a sizeable
portion of the initial laser energy is converted into the energy of
MeV photons ($30$\% in the simulation of Ref.~\onlinecite{Nerush11})
which can have anisotropic distribution in some
cases~\cite{Nerush11_2}.

In this paper we attempt to classify gamma-ray generation regimes and
examine the influence of plasma dynamics, ion acceleration and other
effects on the gamma-ray generation process. For this purpose we
perform a series of numerical simulations for a wide range of foil
densities and laser intensities, as described in Sec.~\ref{sec:map}.
Despite of obvious limitations of such a consideration (e.g., we
restrict the simulations to some certain value of the foil thickness),
it allows us to determine the region of the most efficient gamma-ray
generation (see Sec.~\ref{sec:map}).

At ultrahigh intensities any light nucleus become fully ionized, hence
the ion charge-to-mass ratio doesn't depend much on the material type.
Thus, the presented picture of laser-foil interactions doesn't depend
much on the material type.  Electron dynamics, generation of
electromagnetic fields and ion acceleration are considered in
Sec.~\ref{sec:electrons} and Sec.~\ref{sec:ions}. Though a plenty of
important effects manifest itself in the considered parameter region, it
occurs that gamma-ray generation is strongly affected by ion motion, and
the region of efficient gamma-ray generation approximately coincides
with the region of relativistic ion dynamics (see
Sec.~\ref{sec:ions}).

Having in mind possible applications of MeV photons, we focus on
gamma-ray angular distribution and spectrum along with gamma-ray
generation efficiency.  Characteristics of gamma-ray bunches from
laser-irradiated foils obtained by means of simulations in a wide
range of parameters are discussed in Sec.~\ref{sec:electrons} and
Sec.~\ref{sec:sum}. In Sec.~\ref{sec:sum} two sets of laser-plasma
interaction parameters and the parameters of the resulting gamma-ray
sources are considered in detail. Namely, we discuss the gamma-ray
generation with normally incident $100 \text{ PW}$ laser pulse and
with obliquely incident $3 \text{ PW}$ laser pulse.  In the first case
high laser-to-gamma-ray conversion efficiency ($9\%$) can be easily
got. In the second case tight focusing of the laser pulse and an accurate
choice of the plasma density and the incidence angle let obtain
reasonable conversion efficiency ($\sim 1\%$) along with quite high
directivity of a single-lobe radiation pattern. Possible applications
of the corresponding gamma-ray bunches are discussed and the
comparison with the existing gamma-ray sources is also given.

\section{\label{sec:map} Map of the source gain}

\begin{figure}
\includegraphics{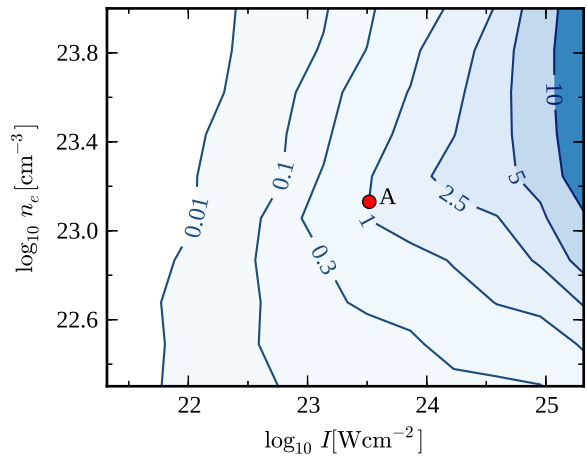}
\caption{\label{fig:gain}The gain $\mathcal{G}$, i.e. the product of the radiation
pattern directivity and the gamma-ray generation efficiency, for
high-energy photons generated at the normal incidence of a $10 \text{
fs}$ linearly polarized laser pulse on a $0.91 \text{ }\upmu \text{m}$
thick plasma slab (see text for further details).}
\end{figure}

Three-dimensional numerical simulations allow us to calculate gamma-ray
radiation pattern, i.e. the directional (angular) dependence of the
emitted energy integrated in time. We introduce the directivity $\mathcal{D}$ of the source as
the ratio of the radiation pattern maximum to its value in
the case of isotropic distribution of the emitted gamma-rays. Since high directivity
of a gamma-ray beam is more reasonable for applications,
here we focus on the gain $\mathcal{G}$ that is the product of the
directivity and the generation efficiency $\mathcal{E}$, where we
introduce $\mathcal{E}$ as the overall
energy of the gamma-rays divided by the initial energy of the laser
pulse. The distribution of the gain on the laser-plasma interaction
parameters is shown in Fig.~\ref{fig:gain}, where interpolation
based on $120$ simulation runs is presented.  Every
computed value corresponds to the normal incidence of a linearly
polarized laser pulse on a fully ionized thin foil. The normalized
amplitude of a laser pulse $a_0 = eE_0/mc\omega$ alters from $25$ to
$2500$, and the initial plasma density normalized to the critical
density $n_0 = n_e/n_{cr}$ lies in the range $15\text{--}745$, where $n_{cr} =
m \omega^2/ 4 \pi e^2$, $e>0$ and $m$ are the magnitude of the
electron charge and the electron mass, respectively, $\omega = 2\pi
c/\lambda$, $\lambda = 0.91 \text{ } \upmu \text{m}$ is the laser
wavelength and $c$ is the speed of light.  The laser pulse has
Gaussian envelope with duration $10 \text{ fs}$ and width $5.4 \text{
} \upmu \text{m}$ (both are measured as FWHM of the laser intensity),
the initial distance between irradiated foil boundary and the laser
pulse centre is $5 \text{ } \lambda$, the interaction time is $12
\lambda/c$, the foil thickness is $l = 1 \lambda$, the ion
mass-to-charge ratio is $M/Z = 2$, where $M$ is the ion mass
normalized to the proton mass and $Z$ is the ion charge normalized to
$e$.

The simulations are performed with a particle-in-cell (PIC) code
that takes into account incoherent emission and decay of hard photons
using the Monte Carlo (MC) technique. The separation of the electromagnetic
fields on a low-frequency (laser field and its harmonics) and
high-frequency (gamma-rays) parts is possible due to a wide gap
between them: spectrum of coherent harmonics emitted in laser-solid
interactions typically extends up to $1 \text{
keV}$~\cite{Brugge10,*Gonoskov11,*Sanz12}, and the characteristic energy of
incoherent photons produced by a laser pulse striking a solid is
greater or of the order of $1 \text{ MeV}$ if laser intensity is
greater than $10^{22} \text{ W} \, \text{cm}^{-2}$ (for
optical wavelengths).

The PIC part of the code that describes laser and plasma fields and
plasma dynamics is based on the same methods as the code
VLPL~\cite{Pukhov99}, however, we use the Vay's particle
pusher~\cite{Vay08} instead of Boris pusher. This part of the code has
been used for a study of various plasma problems, for example, for
simulations of laser-plasma electron acceleration~\cite{Soloviev11}.
The effects of quantum electrodynamics (QED), namely emission of hard
photons and hard photon decay are simulated by an alternative event
generator~\cite{Elkina11} that uses the Baier--Katkov formulae for the
event probabilities~\cite{Baier67,*Baier98,Berestetskii82} which are
applicable in quantum limit (when photon recoil is substantial) as
well as in classical limit. The MC part of the code has been used for
numerical simulations of electromagnetic cascades in a laser
field~\cite{Nerush11,Elkina11}.

\begin{figure}
\includegraphics[width=8.3cm]{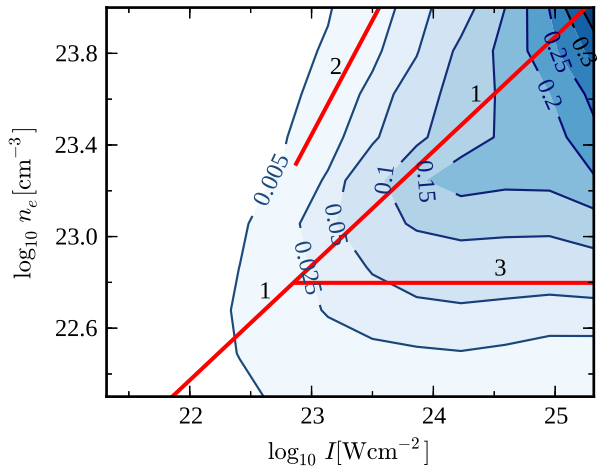}
\caption{\label{fig:phen}The gamma-ray generation efficiency
$\mathcal{E}$ in
laser-foil interaction obtained in the same simulation series as used
for Fig.~\ref{fig:gain}. Lines $1\text{--}3$ show distinctive
interaction regimes. Below line $1$ electrons and ions are easily
separated by the laser field and the laser pulse propagates through
the plasma slab, above this line the plasma reflects the laser pulse.
The region bounded by lines $2\text{--}3$ corresponds to quick foil
acceleration up to relativistic velocity (see Sec.~\ref{sec:ions}).}
\end{figure}

It follows from the simulations that gamma-ray generation is strongly
affected by ion dynamics. The gamma-ray generation efficiency is shown
in Fig.~\ref{fig:phen}, as well as the boundaries of the
characteristic laser-foil interaction regimes. The ions are quickly
accelerated up to relativistic velocities within the region bounded by
lines $2$ and $3$, that fairly well coincides with the region of the
sizeable gamma-ray generation efficiency. The electron dynamics in
laser-foil interactions is discussed in Sec.~\ref{sec:electrons} and
analytical estimations that lead to lines $2$ and $3$ in
Fig.~\ref{fig:phen} are considered in Sec.~\ref{sec:ions}.

\section{\label{sec:electrons}Electron motion in laser-solid interactions}

Electron dynamics in laser-foil interactions can be extremely diverse
and complicated. Nonetheless, as a rule, plasmas prevent laser pulse
penetration and a thin and dense electron layer which screens the
incident field is formed at the front of the laser pulse. As we show
below, this property together with one-dimensional approximation is
enough for a rough description of the electron trajectories.

Dynamics of the electron layer that screens the incident field is
extensively discussed in the context of high-harmonic generation in
laser-solid interactions~\cite{Brugge10,*Gonoskov11,*Sanz12}. It is
shown that at some specific interaction parameters relativistic
electron layers (bunches) with nanometer thickness and density up to
four orders of magnitude higher than nonrelativistic critical plasma
density can be formed and coherently emit high laser harmonics (up to
the photon energy of about $1 \text{ keV}$ that corresponds to the
wavelengths of $1 \text{ nm}$). At the same time a thicker electron
layer, which doesn't emit high harmonics efficiently but also screens
incident field and prevents penetration of the laser pulse into the
plasma, is formed in a very broad range of the interaction parameters.
Despite such a layer doesn't emit keV photons coherently, electrons in
the layer can produce MeV photons due to incoherent synchrotron
emission.

For the description of the electron trajectories we assume that (i)
for normal incidence of a laser pulse on a plasma halfspace in the
framework of one-dimensional geometry a flat thin (much thinner than
laser wavelength) electron layer is formed that (ii) fully compensates
incident laser field behind the layer by layer's fields, hence, the
plasma behind the layer is unperturbed. Additionally we assume that
(iii) ions remain immobile during the interaction and (iv) initial
plasma density is restored behind the layer if it moves towards the
laser pulse. Hence, the surface charge density of the layer is $n_0
x_l$, where $x_l$ is the distance between the initial position of the
plasma boundary and the layer, normalized to $c/\omega$.

The coherent part of the electromagnetic radiation emitted by the
layer at a particular time instance can be easily found from Maxwell's
equations in the reference frame $K'$ where the layer moves along
itself:
\begin{equation}
E'_{\rightarrow} = B'_{\rightarrow} = E'_{\leftarrow} = - B'_{\leftarrow}
= -J'/2,
\end{equation}
where $E$ and $B$ denote the fields of the emitted waves, $y$
component of the electric field and $z$ component of the magnetic
field, respectively, fields are converted from CGS electrostatic units
to the units of $mc\omega/e$, the axes of the $K'$ reference frame are
assumed to be parallel to the axes of the laboratory frame of
reference $K$ in which the linearly polarized laser pulse is incident
along the $x$~axis and the electric field of the pulse directed along
the $y$~axis. The symbols ``$\rightarrow$'' and ``$\leftarrow$''
denote the fields of the waves running in $+x$ and $-x$ directions at
the corresponding layer boundaries, respectively,
\begin{equation}
J' = \int_{x_l'-\delta x_l'/2}^{x_l'+\delta x_l'/2} j'(x') \, dx'
\end{equation}
is the layer surface current density in $K'$, $j$ is the volume current
density in the units of $mc \omega^2/4\pi e $, $x_l'+\delta
x_l'/2$ and $x_l'-\delta x_l'/2$ are the coordinates in $K'$ of the
layer boundaries normalized to $c/\omega$, hence, $\delta x_l'$ is
the layer thickness.  Lorentz transformation of the coherently emitted
fields yields for the laboratory reference frame:
\begin{eqnarray}
E_\rightarrow = B_\rightarrow = -\frac{J}{ 2 ( 1- v_x ) }, \\
E_\leftarrow = -B_\leftarrow = -\frac{J}{ 2 (1+ v_x ) },
\end{eqnarray}
where
\begin{equation}
J = - n_0 x_l v_y,
\end{equation}
$v_x \approx dx_l/dt$ and $v_y$ are the components of the speed of
electrons that form the layer, $x_l$ is the distance between
the layer and the initial (unperturbed) position of the irradiated plasma
boundary in the reference frame $K$, $t$ is the current time instance
normalized to $\omega^{-1}$, and $n_0$ is the initial plasma density
normalized to $n_{cr} = m \omega^2 / 4\pi e^2$. We note again that
$n_0 x_l$ is the surface charge density of the electron layer.

The layer motion can be found now from the assumption that the plasma
remains unperturbed behind the layer because in this region the incident wave
[with $E_y (x,t) = \tilde E(x-t)$, $\tilde E_z(x,t)=0$] is fully
compensated by the wave emitted by the electron layer:
\begin{equation}
\label{eq:epluse}
\tilde E(x_l-t) + E_\rightarrow = 0.
\end{equation}
Assuming that particles in the layer are ultrarelativistic and
$v_x^2 + v_y^2 \approx 1$, the latter equation can be rewritten as
follows:
\begin{equation}
\label{eq:dxldt}
\frac{dx_l}{dt} = \frac{ 4 {\tilde E}^2 (x_l-t) - n_0^2 x_l^2}{ 4
{\tilde E}^2 (x_l-t) + n_0^2 x_l^2}.
\end{equation}

Electrons can leave and join the layer during its motion,
nevertheless, we use Eq.~(\ref{eq:dxldt}) in order to determine some
average trajectory as follows. As $x_l(t)$ is found from
Eq.~(\ref{eq:dxldt}), then $v_x(t)$ and $v_y = \pm \sqrt{1- v_x^2}$
can be obtained, and sign of $v_y$ is chosen to satisfy
Eq.~(\ref{eq:epluse}).

Eq.~(\ref{eq:dxldt}) is not based on the equations of electron
motion, and it should not be affected much by such effects as radiation losses and back reaction
of the coherently emitted fields on the layer. Hence, the main
limitation of this equation is the assumption of immobile ions.

Due to relativistic motion of the layer $E_\rightarrow \neq
E_\leftarrow$, hence, the amplitude and the shape of the reflected laser light can be
considerably modified in comparison with the incident light. This also
means that the fields of the incident laser pulse are not compensated
inside the electron layer. However, the Lorentz factor of the layer
electrons cannot be estimated in the framework of the proposed model
because the model leads to the following:
\begin{equation}
\frac{d \gamma}{dt} = - \mathbf v \mathbf E \approx -v_x
E_x - v_y \left( \tilde E + E_\leftarrow \right) = 0,
\end{equation}
where we assume that $E_x = n_0 x_l$. Hence, energy gain of the
electrons is caused by more subtle effects such as finiteness of the
gamma-factor, dispersion of electron parameters, electron inertia, ion
motion etc. that hardly could be taken into account. Since electron
gamma-factor is the crucial parameter for incoherent photon emission,
gamma-ray generation cannot be described only on the basis of Eq.~(\ref{eq:dxldt}).
Nevertheless, Eq.~(\ref{eq:dxldt}) allows us to estimate the depth on
that laser pulse pushes the electron layer.

Maximal value of the layer displacement $X_l$ corresponds to $dx_l/dt
= 0$.  Assuming that the maximal displacement corresponds also to the
maximal value of the incident field $\tilde E \approx a_0$, we obtain:
\begin{equation}
\label{eq:Xl}
X_l \simeq \frac{2}{S}, \qquad S = \frac{n_0}{a_0},
\end{equation}
where $S$ is the so-called similarity parameter~\cite{Gordienko05}.
Eq.~(\ref{eq:Xl}) is very useful for a qualitative description of
the ion dynamics, as shown in the next section.
Moreover, Eqs.~(\ref{eq:dxldt}) and (\ref{eq:Xl}) can be used in order
to analyse gamma-ray parameters as follows.

\begin{figure}
\includegraphics[width=8.3cm]{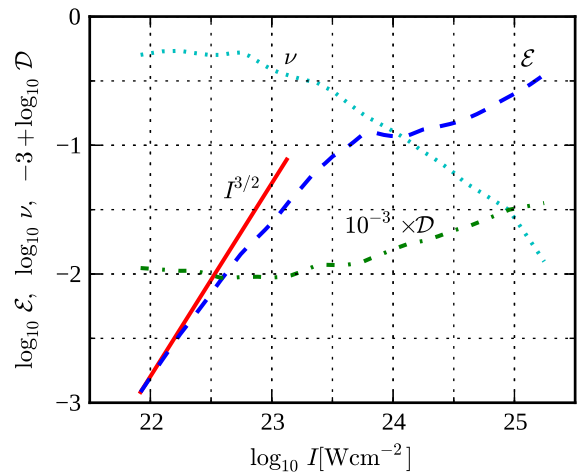}
\caption{\label{fig:section}(Dashed line) The gamma-ray generation efficiency
$\mathcal{E}$, (dotted line) the spectral coefficient $\nu$, (dashed-dotted line) the directivity
$\mathcal{D}$ at fixed value of the similarity
parameter $S = 2/l = 1/\pi$ (i.e., along line $1$ in Figs.~\ref{fig:phen}
and~\ref{fig:ions}) and (solid line) the curve $\mathcal{E} \propto
I^{3/2}$ versus laser intensity. Results of numerical simulations for
the same interaction parameters as used for Fig.~\ref{fig:gain}.}
\end{figure}

For the maximal number of electrons in the layer
Eq.~(\ref{eq:Xl}) yields $N \propto n_0 X_l \sim a_0$. It also follows
from the considered model that the shape of the electron trajectories,
hence, the curvature radius, depends only on the
similarity parameter $S$. Since synchrotron emitting power
proportional to $\gamma^4$ for the fixed curvature radius,
for fixed $S$ the gamma-ray generation efficiency scales as $\mathcal{E} \propto N
\gamma^4/I \propto
I^{3/2}$ under the assumption that the average electron Lorentz factor $\gamma
\sim a_0$. The dependence of the gamma-ray generation efficiency on
laser intensity at fixed $S$ is shown in Fig.~\ref{fig:section} along
with the fit $\mathcal{ E} \propto I^{3/2}$ that fairly well coincides
with numerical data at low intensities.

The dependence of the directivity $\mathcal D$ and spectral
coefficient $\nu$ on intensity is also shown in
Fig.~\ref{fig:section}. The coefficient $\nu$ is determined by
the root-mean-square fitting in the logarithmic axes of the obtained gamma-ray
spectrum by the classical synchrotron spectrum of a single
electron~\cite{Landau75}:
\begin{eqnarray}
\label{eq:nu} \frac{dN_{ph}}{d\omega_{ph}} & \propto & \int_\kappa^\infty
\operatorname{Ai}(\xi) \, d \xi + \frac{2}{\kappa}
\operatorname{Ai}'(\kappa), \\
\kappa & = & \left( \frac{\omega_{ph}}{\nu a_0^3 \omega}
\right)^{2/3},
\end{eqnarray}
where $\hbar \omega_{ph}$ is hard-photon energy.  Actually in the
classical synchrotron spectrum $\kappa^{3/2} = \omega_{ph} / F_\perp
\gamma^2 \omega$, where $F_\perp$ is the perpendicular to the electron
velocity component of the force that causes photon emission,
normalized to $mc\omega$.  Obviously, we set $F_\perp \gamma^2 = \nu
a_0^3$, hence, $\nu$ can characterize such effects as decrease of the
electron Lorentz factor and decrease of the angle between $\mathbf{F}$
and the electron velocity caused by the radiation
reaction~\cite{Fedotov10}.  We found that if the appropriate spectral
coefficient $\nu$ is chosen, Eq.~(\ref{eq:nu}) describes well enough
the part of the gamma-ray spectrum extended from the gamma-ray energy cut-off
to $0.2$ of its value.

It should be noted that for $I\lesssim 10^{23} \text{
W}\,\text{cm}^{-2}$ the directivity and the spectral coefficient do not change with the
intensity (see Fig.~\ref{fig:section}), that conforms to the proposed analytical model. Besides,
for $I\gtrsim 10^{23} \text{ W}\,\text{cm}^{-2}$, $\nu$ declines
apparently due to radiation losses and $\mathcal{D}$ increases with
the increase of the intensity. The latter
can be interpreted as the radiation pattern narrowing due to the light
aberration and the relativistic Doppler effect which are caused in turn by
relativistic motion of the entire foil. This effect is also considered
in Ref.~\onlinecite{Ridgers12}.

\begin{figure}
\includegraphics[width=8.3cm]{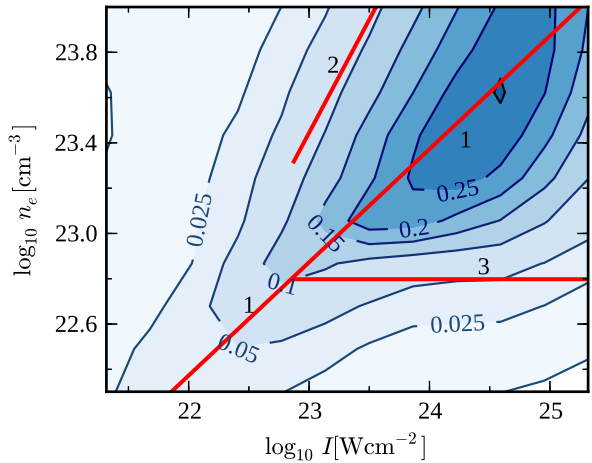}
\caption{\label{fig:ions}The overall ion energy
normalized to the initial energy of the laser pulse for the same
set-up as in Fig.~\ref{fig:gain}. Lines $1$, $2$ and $3$ are the same
as in Fig.~\ref{fig:phen}.}
\end{figure}

\section{\label{sec:ions}Ion dynamics in laser-foil interactions}

Regimes of laser-foil interactions in terms of ion motion can be
classified by means of Eq.~(\ref{eq:Xl}). First, the line
\begin{equation}
\label{eq:2l}
S = \frac{2}{l},
\end{equation}
where $l$ is the foil thickness, obviously, corresponds to a complete
separation of the electrons and the ions at the maximal magnitude of the laser
intensity. Under the assumption of immobile ions, for $S>2/l$
the rear side of the foil remains unperturbed, otherwise,
for $S<2/l$, almost all electrons are expelled from the foil.
Eq.~(\ref{eq:2l}) corresponds to line $1$ in Figs.~\ref{fig:phen}
and \ref{fig:ions}.

The magnitude of the electric field that accelerates the ions
for $S>2/l$ can be
estimated as follows:
\begin{equation}
E_x \simeq n_0 X_l = 2 a_0.
\end{equation}
Let us introduce the time interval $\tau$ such that ions are accelerated by $E_x$ up
to relativistic velocities during this interval. Hence, $2 \tau
a_{0,i} \approx 1$, where
\begin{equation}
a_{0,i} = \frac{m}{m_p} \frac{Z}{M} a_0.
\end{equation}
Here $m_p \approx 1836 m$ is the proton mass.
At the same time, relativistic ions pass the distance $X_l$ during
the time $X_l$, equating that to $\tau$ yields the following
relation between $a_0$ and $n_0$:
\begin{equation}
\label{eq:n0top}
n_0 \simeq 4 a_0 a_{0,i}.
\end{equation}
Thus, if the plasma density is higher than that given by Eq.~(\ref{eq:n0top}),
$X_l$ is small and the ions have time to leave the accelerating gap
before they gain relativistic velocities. In the opposite case, if the
plasma density is lower than the threshold Eq.~(\ref{eq:n0top}),
accelerating gap is thick enough and the ions at the front of the foil
become relativistic. In the latter case the foil is crumpled and blown
away by the laser pulse with relativistic velocity until the density
of the shovelled plasma is not high enough to slow down the process.
Eq.~(\ref{eq:n0top}) corresponds to line $2$ in Figs.~\ref{fig:phen} and \ref{fig:ions}.
It is worth to note that the estimations derived above relates mostly to
high enough laser intensities ($a_{0,i} \gtrsim 1/4 \pi$).  Otherwise
$\tau$ becomes greater than the laser period and the ion acceleration
process depends on the laser pulse duration.

The next threshold that corresponds to relativistic ion dynamics can
be found in the region $S<2/l$. The completely separated electrons and
ions generate in this case the following accelerating field:
\begin{equation}
E_x \simeq n_0 l.
\end{equation}
This field accelerates the ions up to relativistic velocities during
the laser period if $n_0 > \hat n_0$, where
\begin{equation}
\label{eq:n0bottom}
\hat n_0 = \frac{1}{2 \pi l} \frac{m_p}{m} \frac{M}{Z}.
\end{equation}
Despite ions on the front side of the foil are irradiated, the direct
acceleration by the laser pulse is weaker than the acceleration by the induced plasma
fields and can be neglected~\cite{Esirkepov04}. The threshold $n_0 =
\hat n_0$ is shown in Figs.~\ref{fig:phen} and \ref{fig:ions} as line $3$.

\begin{figure}
\includegraphics[width=8.3cm]{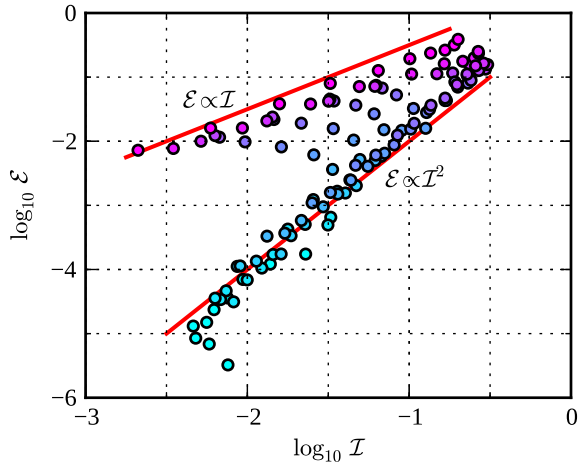}
\caption{\label{fig:astro}The gamma-ray generation efficiency as
function of the overall ion energy normalized to the initial energy of the
laser pulse. The circles correspond to the results of numerical simulations
used for Fig.~\ref{fig:phen}. The colour corresponds to the
laser intensity, from (cyan) $2\times10^{21} \text{
W}\,\text{cm}^{-2}$ to (magenta) $2\times10^{25} \text{ W}\,
\text{cm}^{-2}$.}
\end{figure}

Summarizing, on the $(\text{intensity}, \text{ density})$ plane the
region characterized by quick ion acceleration up to
relativistic velocities is bound between lines corresponding to
Eqs.~(\ref{eq:n0top}) and (\ref{eq:n0bottom}) (lines $2$ and $3$ in
Figs.~\ref{fig:phen} and \ref{fig:ions}). Surprisingly, this region
approximately coincides with the area of efficient gamma-ray
generation, as seen from Figs.~\ref{fig:phen} and \ref{fig:ions}.
Besides, plasma dynamics is diverse enough in this region, e.g., it is
quite irregular if $S>2/l$ and regular if $S<2/l$, at high intensities
it is significantly affected by abundant positron production, etc. The
considered coincidence can be partially explained by the proximity of
the threshold intensity value for fast ion acceleration ($a_{0,i} =
1/4 \pi$, that yields $I \sim 10^{23} \text{ W}\,\text{cm} ^{-2}$) and
the threshold intensity value for the radiation-dominated regime.
Furthermore, two distinctive regimes with an evident relation between ion
and gamma-ray energies are revealed by the results of the numerical simulations
(see Fig.~\ref{fig:astro}). Independently on the plasma density, at low
laser intensity the final gamma-ray energy scales as the squared overall ion
energy, and at high intensity the gamma-ray energy is in direct ratio with
the ion energy. Thus, the mutual influence of gamma-ray generation and
ion acceleration requires additional investigations.

\begin{figure}
\includegraphics[width=8.3cm]{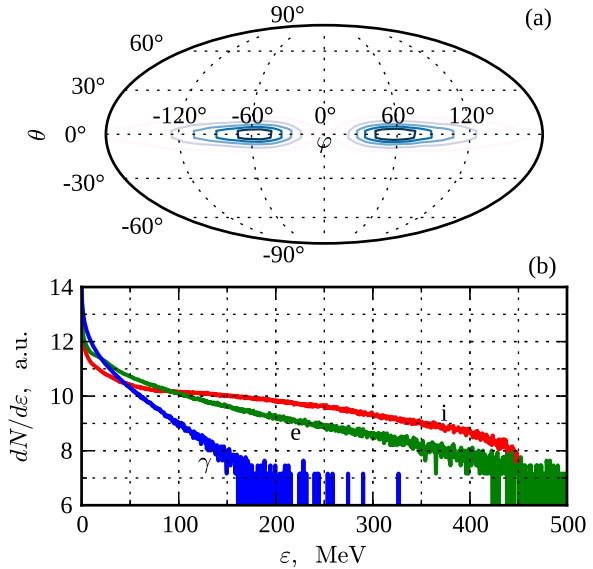}
\caption{\label{fig:A}(a) The Mollweide projection of the gamma-ray
radiation pattern and (b) the photon, electron and ion spectra
obtained in the numerical simulation corresponding to the point ``A'' in
Fig.~\ref{fig:gain}. The longitude $\varphi$ counts in the
polarization plane, and the latitude $\theta$ counts from the
polarization
plane so that point $\varphi=0$, $\theta=0$ corresponds to the initial
propagation direction of the laser pulse.}
\end{figure}

\section{\label{sec:sum}Summary and discussion}

In conclusion, incoherent generation of hard photons in normal
incidence of a laser pulse on a foil is considered by means of
three-dimensional PIC+MC simulations. For various plasma densities and
laser intensities the gamma-ray gain, directivity, generation efficiency
and spectral features are found. The influence of ion dynamics on
the emission of gamma-rays and the influence of radiation losses on
the electron dynamics are discussed.

As seen from Fig.~\ref{fig:gain}, the maximal gain at the intensity
$3.2 \times 10^{23} \text{ W}\, \text{cm}^{-2}$ ($a_0 = 310$) can be
obtained if foil with $n_e = 1.1 \times 10^{23} \text{ cm}^{-3}$ ($n_0
= 85$) is used (see point ``A'' in Fig.~\ref{fig:gain}). This
intensity level is potentially attainable in the nearest future, so,
let us consider in detail some properties of the gamma-ray bunches
produced at this point.  Thus, the generation efficiency is $9\%$,
the directivity is $\mathcal{D} \approx 12$, hence, the gain is about
unity in this case. It should be noted that the generation efficiency
drops dramatically at lower intensities and saturates at higher
intensities, as seen from Fig.~\ref{fig:section}. The Mollweide
projection of the gamma-ray radiation pattern and the photon spectrum
are shown in Fig.~\ref{fig:A}. Two-lobe radiation pattern with lobes
confined to the polarization plane is typical for the case of normal
incidence~\cite{Nakamura12,Ridgers12}. One-lobe radiation pattern
directed towards the incident laser pulse analogous to that in
Ref.~\onlinecite{Brady12} reveals itself only at quite low intensities
when the generation efficiency is not high.

The gamma-ray source corresponding to the point ``A'' in
Fig.~\ref{fig:gain} for $1 \text{ Hz}$ laser shot frequency produces
every second $5 \times 10^{13}$ photons with energy greater than $0.5
\text{ MeV}$. For the photon energy of $1 \text{ MeV}$ such a source
provides the spectral power of $2 \times 10^{10} \text{ photons}/
\text{s}/ 0.1\% \text{ bandwidth}$ and the spectral intensity of $2 \times
10^4 \text{ photons}/ \text{mrad}^2/ \text{s}/ 0.1\% \text{ bandwidth}
$. Now let us compare these parameters with the parameters of modern
Compton gamma-ray sources and discuss the potential applications of
such a laser-plasma synchrotron source.

One of the most intense gamma-ray sources, Hi$\gamma$S~\cite{Weller09},
in a high-flux mode can produce about $10^9 \text{ photons}/\text{s}$
with average energy of $1 \text{ MeV}$. A collimator that provides
$3\%$ photon energy spread reduces the flux by the order of magnitude.
Hence, HI$\gamma$S on the level of $1 \text{ MeV}$ provides the spectral
power of $3 \times 10^6 \text{ photons}/\text{s}/0.1\% \text{
bandwidth}$ that is four orders of magnitude lower than that for the
considered laser-plasma source. However, HI$\gamma$S's gamma-ray beams
after the collimator spread by the angle of about $20 \text{ mrad}$ that
yields the spectral intensity of $10^4 \text{ photons}/ \text{mrad}^2/
\text{s}/ 0.1\% \text{ bandwidth}$ comparable with the considered
all-optical gamma-ray source.

Another Compton source, produced quasimonoenergetic photon beams with
the energy ranging from $0.1$ to $0.9 \text{ MeV}$, has been recently
developed~\cite{Gibson10} at LLNL and used to perform nuclear resonance fluorescence
(NRF) experiments~\cite{Albert10}. This source yields $10^5
\text{ photons}/\text{s}$ with average energy corresponding to $^7
\text{Li}$ line ($478 \text{ keV}$) and energy spread of $12\%$.
The resulting spectral power of $10^3 \text{ photons}/\text{s}/0.1\%
\text{ bandwidth}$ is enough for the accurate detection of $^7
\text{Li}$ isotope \textit{in situ}, however this required $7.5$ hours
of the operation~\cite{Albert10}. The divergence of this gamma-ray beam is
about $10 \text{ mrad}$, hence, not only the spectral power, but also the
spectral intensity of the LLNL source is much lower than that for the
considered laser-plasma source.

Despite of high potential performance of the laser-plasma gamma-ray
sources, some properties of them are so unfamiliar for nowadays
gamma-ray physics that more sophisticated NRF experimental techniques
can be required.  For instance, the inherently wide photon spectrum will
lead to extremely high Compton background in spectral measurements of
re-emitted photons.  Moreover, semiconductor and scintillating
detectors generally used for gamma-ray energy measurements can operate
only at the flux less than one photon per nanosecond. Thus, the NRF
experiments will require detector arrays and a lot of the laser shots in
order to obtain reasonable statistics.  Nevertheless, femtosecond
duration of gamma-ray bunches from laser-irradiated plasmas together
with relatively long lifetimes of low-laying nuclear levels (generally
from tens of femtoseconds to picoseconds~\cite{Chadwick06}) might enable the
time-of-flight separation~\cite{Ahmed04} of the Compton and NRF signals if
an appropriate experimental design is proposed.

Obviously, gamma-ray beams from laser-plasma sources can be used in a
number of applications that do not require high spectral quality. One
of such an application is the radiography of ultrahigh density matter.
It can be performed by means of laser-plasma sources with
unprecedented time resolution, that is crucial for fast ignition
experiments~\cite{Barty04}. Another promising application of
laser-plasma gamma-ray sources is a high-flux positron source based on
pair creation by high-energy photons in a material target. Proposals
of International Linear Collider suppose not only high electron and
positron energy, but also high luminosity, that requires a
reinforcement of nowadays positron sources.  Relatively efficient
positron production using Compton gamma-ray sources now is proved in
the experiments, where gamma-to-positron conversion efficiency of
about $10^{-3}$ and total positron flux of $10^4 /\text{s}$ are
achieved~\cite{Omori06}. Since the conversion efficiency does not
sharply depend on the gamma-ray energy, the average flux of about $10^{10}
\text{ positrons}/\text{s}$ can be expected if the considered
laser-plasma gamma-ray source is used.

Keeping on speculating about possible future experiments with bright
gamma-ray beams, propagation-based polychromatic phase-contrast
imaging~\cite{Wilkins96} can be proposed. According to van
Cittert--Zernike theorem the scale of the spatial coherence of light
emitted by an incoherent source is about the product of light
wavelength, distance to the source and reverse size of the source.
Hence, the scale of the spatial coherence of a laser-plasma gamma-ray
source is about $10 \text{ } \upmu \text{m}$, if the photon energy is
$1 \text{ MeV}$, the source size is $1 \text{ } \upmu \text{m}$ and
the distance from the source is $10 \text{  m}$.  Assuming that $100$
gamma-photons belong to the corresponding solid angle, at least
$10^{8} \text{ photons}/\text{mrad}^2$ are required for the
phase-contrast gamma-imaging, that nearly the value of the considered
source. Matter refractive index drops rapidly with the decrease of a
hard-photon wavelength, however, the refractive index of polarized
vacuum does not depend on the wavelength~\cite{Berestetskii82}. Hence,
for a photon beam propagating through polarized vacuum the induced
curvature of the beam phase front is higher for smaller wavelengths.
This fact along with femtosecond duration of gamma-ray bunches from
laser-irradiated plasmas can make such bunches a promising probe for
the experiments on vacuum polarization in ultraintense laser field
proposed in Ref.~\onlinecite{King10,*King10_1}.

Finally let us note that modern high-power laser facilities are
already appropriate for quite efficient generation of high-energy
photons in laser-plasma interactions. Simulation results of oblique
incidence (the incidence angle is $60 ^\circ$) of a tightly focused $3
\text{ PW}$ $p$-polarised laser pulse (FWHM: $3 \text{ }\upmu
\text{m}$, $10 \text{ fs}$; $\lambda=0.91 \text{ }\upmu \text{m}$,
$a_0=100$, the peak intensity is $3.3 \times 10^{22} \text{ W}\,
\text{cm}^{-2}$) on a foil ($n_e=1.7 \times 10^{23}
\text{ cm}^{-3}$, $M/Z=2$, $l=1 \lambda$) demonstrates reasonable
generation efficiency ($0.7\%$) and unexpectedly very high directivity
($\mathcal{D} = 37$). The radiation pattern in this case consist of only
one lobe lying both in the plane of incidence and in the foil plane,
hence, the angle between the initial propagation direction of the laser
pulse and the lobe direction is $30^\circ$. Under the assumption of $1
\text{ Hz}$ laser system the resulting total gamma-ray flux is
$10^{12} \text{ photons}/\text{s}$, the spectral power is $10^8 \text{
photons}/ \text{s} / 0.1\% \text{ bandwidth}$ and the spectral intensity
is $400 \text{ photons}/ \text{mrad}^2/ \text{s}/
0.1\%\text{bandwidth}$ at the energy level of $1 \text{ MeV}$. Thus,
quite high flux, spectral power and directivity of such a source can
make it desirable for a number of applications and experiments.

Summarizing, theoretical analysis and three-dimensional PIC+MC
simulations demonstrate an interplay of ion acceleration and hard
photon generation in laser-foil interactions. The scaling of the
gamma-ray bunch parameters with the laser intensity is discussed.
Properties and possible applications of the resulting gamma-ray
bunches are considered, including nuclear resonance fluorescence and
high-flux positron sources.

\begin{acknowledgments}
This work has been supported in part by the Government of the Russian
Federation (Project No. 14.B25.31.0008) and by the Russian Foundation
for Basic Research (Grants No  13-02-00886, 13-02-97025).
\end{acknowledgments}

\bibliography{main}
\end{document}